\documentclass[12pt]{article}
\usepackage{amsmath,amsthm,url,graphicx}

\newcommand{\I}{\makebox[0pt][l]{1}\hspace{0.3ex}\mbox{1}}
\newcommand{\squash}[1]{\raisebox{0.04ex}[0pt][0pt]{\small$\textstyle #1$}}

\newcommand{\ket}[1]{\vert#1\rangle}
\newcommand{\bra}[1]{\langle#1\vert}
\newcommand{\Tr}{\text{Tr}}
\newtheorem{theo}{Theorem}
\theoremstyle{definition}
\newtheorem{lemma}{Lemma}
\newtheorem{corol}{Corollary}

\begin{document}

\title{Minimum entangled state dimension\\required for pseudo-telepathy}

\author{Gilles Brassard\thanks{Supported in part by
the Natural Sciences and Engineering Research Council of Canada ({\sc Nserc}),
the Canadian Institute for Advanced Research ({\sc Ciar}),
the Mathematics of Information Technology and Complex Systems Network ({\sc Mitacs})
and the Canada Research Chair Programme.}
\hspace{0.75cm} Andr\'e Allan M\'ethot \hspace{0.75cm} Alain
Tapp\thanks{Supported in part by {\sc Nserc}, {\sc Ciar}, {\sc Mitacs} and Qu\'ebec's {\sc Fqrnt}.}\\[0.5cm]
\normalsize\sl D\'epartement d'informatique et de recherche op\'erationnelle\\[-0.1cm]
\normalsize\sl Universit\'e de Montr\'eal, C.P.~6128, Succ.\ Centre-Ville\\[-0.1cm]
\normalsize\sl Montr\'eal (QC), H3C 3J7~~\textsc{Canada}\\
\normalsize\texttt{\{brassard,\,methotan,\,tappa\}}\textsf{@}\texttt{iro.umontreal.ca}
}

\date{16 December 2004}

\maketitle
\begin{abstract}
Pseudo-telepathy provides an intuitive way of looking at Bell's inequalities,
in which it is often obvious that feats achievable by use of
quantum entanglement would be classically impossible.
A~two-player pseudo-telepathy game proceeds as follows:
Alice and Bob are individually asked a question
and they must provide an answer.
They are \emph{not} allowed any form of
communication once the questions are asked, but they may
have agreed on a common strategy prior to the execution of the game.
We~say that they \textit{win} the game if the questions and answers
fulfil a specific relation. A~game exhibits \textit{pseudo-telepathy}
if there is a quantum strategy that makes Alice and
Bob win the game for all possible questions, provided
they share prior entanglement, whereas it would be impossible to win
this game systematically in a classical setting.
In~this paper, we show that any two-player
pseudo-telepathy game requires the quantum players to share
an entangled quantum system of dimension at least~\mbox{$3 \times 3$}.
This is optimal for two-player games, but the most efficient
pseudo-telepathy game possible, in terms of total dimension, involves
\textit{three} players who share a quantum system of
dimension~\mbox{$2 \times 2 \times 2$}.
\end{abstract}

\section{Introduction}

A two-player \emph{game} $G$
is a tuple $(X,Y,A,B,R)$, where
$X$, $Y$\!, $A$ and $B$ are finite sets and
\mbox{$R \subseteq X \times Y \times A \times B$} is a relation amongst those sets.
In an \textit{instance} of the game, one player, Alice, is asked a question
$x \in X$ and she must produce an answer $a \in A$. 
The~other player, Bob, is asked
a question $y \in Y$ and he must produce an answer $b \in B$.
The players are not allowed to communicate after they have
received their questions.
Alice and Bob \textit{win} the instance if \mbox{$(x,y,a,b) \in R$}\,;
they have a \textit{winning strategy} if they
can win systematically every instance. 
Note that these games are sometimes introduced in the literature with the addition
of a \textit{promise} that the questions must fulfil, but we
can ignore this notion without loss of generality because promises
can always be worked inside the relation~$R$.
If~the question \mbox{$(x,y) \in X \times Y$} should not be asked because
it does not fulfil the promise, simply add \mbox{$(x,y,a,b)$} to $R$ for
all possible \mbox{$a \in A$} and~\mbox{$b \in B$}.

We say that $G$ is a \emph{pseudo-telepathy game of
dimension $d_A \times d_B$} if it does not have a \mbox{classical}
winning strategy, yet a quantum winning strategy exists, provided
the players share a prior entangled state of dimension \mbox{$d_A \times d_B$}.
The term ``pseudo-telepathy'' was coined for this phenomenon because
it corresponds to a \mbox{behaviour} that cannot be explained without \textit{some} form
of communication in any classical local realistic world.
Imagine classical \mbox{physicists} who observe this phenomenon.
Imagine further that they have placed the two players in space-like
separated regions by asking questions with enough simultaneity
and requesting answers so quickly that signals sent at the speed of
light by either player would arrive too late to inform the other
player before answers must be produced.
The fact that the players continue to answer correctly every time
\textit{even though this is classically impossible}---or~at least
overwhelmingly unlikely---%
given that they cannot communicate by any method known to a classical
physicist would be puzzling.  So~puzzling in fact that the only ``\mbox{reasonable}''
expla\-na\-tion would be that the players communicate (since they must!)\ by
ways yet unknown to physics.
Well, why not telepathy then?
Furthermore, what better proof that telepathic communication
must be superluminal!
(Of~course, the correct explanation is quantum \mbox{mechanics}, not telepathy.)
Please read~\cite{BBT04a} for a survey of pseudo-telepathy games.
It~is easy to extend the concept of pseudo-telepathy to more than two players~\cite{BBT04b}.

There is a direct connection between pseudo-telepathy
and Bell's Theorem~\cite{bell64,bc90,CHSH,cirelson80}
since John Bell gave the first proof that some bipartite phenomena
can be observed quantum-mechanically with joint probabilities that
would be impossible between
classical systems that do not communicate.
Later work on so-called ``Bell inequalities without inequalities'' (sic!)\,%
\footnote{\,This terminology is unfortunate, not only because it sounds---and~is!---silly, also because
the only \mbox{\textit{inequality}} involved
here is that zero is smaller than any nonzero positive number!}
or ``Bell inequalities without probabilities''~%
\cite{aravind99,cabello01a,ghz89,hardy92,mermin90a}
are even more relevant to pseudo-telepathy.
Recall that ``Bell's theorem'' is the name usually given to an inequality, or set of
inequalities, that the expectation values of the outcomes to a
bipartite measurement have to respect according to any classical local realistic theory,
but that are violated by quantum mechanics.
A~``Bell inequality without probabilities'' consists in a similar set of measurements and
measurement outcomes, except that it is sufficient to concentrate on which outcomes
are possible and which are not possible according to quantum mechanics to reach
a situation that would be impossible according to any classical local realistic theory. 
The questions raised by these theorems and inequalities are at the core of
many discussions on the nature of our world and on the
interpretation of quantum mechanics.

A~consequence of this paper is that, despite obvious similarities,
pseudo-telepathy is a notion \textit{strictly stronger} than
Bell inequalities without probabilities. Indeed,
non-maximally entangled two-qubit states, such as Lucien Hardy's
\mbox{$\ket{\Gamma}=\squash{\frac{1}{\sqrt{3}}}\bigl(\ket{01}+\ket{10}+\ket{11}\bigr)$},
give rise to~nonclassical correlations~\cite{hardy92} when each qubit is measured
independently at random either in the computational or in the Hadamard basis.
This classical impossibility remains even if we consider only which joint
measurement outcomes can or cannot happen, \mbox{ignoring} the specific probabilities.
Nevertheless, we prove in this paper that there is no way to turn Hardy's
\mbox{correlations} into a pseudo-telepathy game because
Hardy's two-qubit state $\ket{\Gamma}$ \mbox{provides} enough entanglement for the
emergence of a Bell inequality without probabilities,
but not enough for pseudo-telepathy.
On~the other hand, it is straightforward to transform any pseudo-telepathy game
into a Bell inequality without probabilities.

After this Introduction,
we present a brief history of pseudo-telepathy games in
Section~\ref{history} and we state our main result:
there cannot exist a pseudo-telepathy game of dimension~\mbox{$2 \times 2$}.
It~follows that the smallest two-player pseudo-telepathy game requires a
shared entangled state of dimension at least~\mbox{$3 \times 3$},
and therefore the optimal total dimension belongs to a
three-player game known since 1990, in which the players share
an entangled state of dimension~\mbox{$2 \times 2 \times 2$}.
Section~\ref{sharpen} reviews the main tools that we use, such as
the notion of generalized measurements (POVMs), and sharpens these
tools to make them more appropriate for our purpose.
Section~\ref{main}
proves our main result and its corollaries.
Finally, we conclude and propose an intriguing open question in Section~\ref{concl}.

\section{History of pseudo-telepathy and statement of result}\label{history}

The history of pseudo-telepathy can be traced back to 1983, when
Peter Heywood and Michael Redhead~\cite{hr83} discovered a way to combine entanglement
with the noncontextuality theorem of Simon Kochen and Ernst Specker~\cite{KS67},
allowing them to propose an experimentally testable version of the
Kochen-Specker theorem. (The original Kochen-Specker theorem was inherently
counterfactual, and therefore not amenable to experimental verification.)
Even though they did not express their idea in those terms, the approach
of Heywood and Redhead was reinterpreted as a Bell inequality without probabilities
fifteen years later by Padmanabhan Aravind~\cite{aravind99}, and eventually as an explicit pseudo-telepathy
game by Richard Cleve, Peter H{\o}yer, Ben Toner and John Watrous (CHTW)~\cite{chtw04} in~2004.

The prior art of ``Bell inequalities without probabilities'' was introduced in 1989
by Daniel Greenberger, Michael Horne and Anton Zeilinger (GHZ)
in a four-party scenario~\cite{ghz89}.
This breakthrough was simplified to a \textit{three-party} scenario and greatly popularized
by David Mermin~\cite{mermin90a}.
It~has been argued
that Mermin's 1990 paper provided the first \textit{explicit}
pseudo-telepathy game~\cite{QCCsurvey}, although the term ``pseudo-telepathy''
had yet to be invented.

The notion of pseudo-telepathy was formalized in 1999 (not yet under that name)
by Gilles Brassard, Richard Cleve and Alain Tapp~\cite{bct99}. 
They gave the first explicit \textit{two-player} game,
for which they showed that pseudo-telepathy of dimension $n \times n$
occurs for all sufficiently large~$n$.
Later, Viktor Galliard, Stefan Wolf and Alain Tapp~\cite{gwt02}
proved that the specific value $n=16$ gives rise to
pseudo-telepathy in that game.
In~the mean time, other
two-player pseudo-telepathy games
had been discovered that required a smaller dimension,
such as Aravind's \textit{magic square}~\cite{aravind02}
and equivalent games~\cite{cabello01a,cabello01b}
of dimension \mbox{$4 \times 4$}.

Could a pseudo-telepathy game of smaller dimension exist?
The answer came in 2004 when CHTW~\cite{chtw04}
reinterpreted the original result of Heywood and Redhead~\cite{hr83}, as
mentioned above, into a pseudo-telepathy game of
dimension \mbox{$3\times 3$}, in which the output sets $A$ and $B$
are of cardinality~2 and~3, respectively. 
In~the same paper, they proved that the output sets in \textit{any}
two-player
pseudo-telepathy game cannot both be of cardinality as small as~2,
which established the optimality of their game according to that criterion.
But was their game also optimal in terms of the dimension of the shared prior
entanglement?

A von Neumann (projective) measurement on a quantum system of dimension $d$
cannot produce more than $d$ distinct outputs.  It~follows from the minimum
size of output sets in any pseudo-telepathy game~\cite{hr83} that a game of dimension
$2 \times 2$ cannot exist if the players are restricted to measuring their
share of the prior entanglement with a von Neumann measurement (without
the help of ancillary quantum systems). This raises a natural question:
Could a pseudo-telepathy game of dimension $2 \times 2$ exist if the players
are allowed to perform generalized measurements (POVMs---see~Section~\ref{sharpen}) on their quantum systems?
We~answer this question by the negative: both $d_A$ and $d_B$ must be at least~$3$
for a two-player pseudo-telepathy game of dimension $d_A \times d_B$ to exist.
It~follows that the game of CHTW~\cite{chtw04} is optimal also in terms of dimension
among all possible two-player pseudo-telepathy games.
If~we allow more participants, however, the older game of GHZ/Mermin~\cite{ghz89,mermin90a}
is optimal in terms of total quantum dimension.

\section{Sharpening the quantum tools}\label{sharpen}

Let us first review the notion of generalized measurements, also known as
Positive Operator Valued Measures, or POVMs for short~\cite{asher}, state some known results,
and then sharpen the tools for our purpose.
POVMs are the most general type of measurement allowed by quantum mechanics.
They are described by a collection of POVM \emph{elements}.
Each POVM element is a positive matrix~$M_i$, i.e.~a matrix of the form
$M_i= D_i^{\dagger}D_i$ for some matrix $D_i$.
The collection $\{M_i\}$ forms a POVM under the condition that
$\sum_i M_i = \I$, the identity matrix.  When applied
on state $\rho$, each possible value $i$ is produced as the classical
outcome of the POVM
with probability $\Pr[i]=\Tr(\rho M_i)$.
In~case $\rho=\ket{\Psi}\!\bra{\Psi}$ is a pure state, this probability can
be written equivalently as $\Pr[i]=\bra{\Psi} M_i \ket{\Psi}$.
In~general, there could be a quantum state leftover in addition to the
classical outcome, but this is irrelevant for the purpose of pseudo-telepathy.

Generalized measurements are justified as a physical process by \textit{Naimark's Theorem}
(sometimes transliterated from the Russian as Neumark's Theorem): POVMs are equivalent to
adding an ancillary quantum system in a known state to the state under measurement,
and then performing an ordinary von Neumann projective measurement on the joint
quantum system.

Even though POVM elements can be arbitrary positive matrices,
our main result is easier to derive if we restrict them to be proportional to projection operators.
The next lemma (due to~\cite{cgm99})
establishes that this simplification can be taken without loss of generality.

\begin{lemma}
\label{sd}
Any POVM can be rewritten in such a way that all its elements are
proportional to one-dimensional projectors.
\end{lemma}

\begin{proof}
Consider a POVM whose elements form the collection~$\{M_i\}$.
From the spectral decomposition theorem,
each of the $M_i$ can be written as $M_i=\sum_j b_{ij}P_{ij}$,
where the $b_{ij}$ are real constants,
$0 <  b_{ij} \le 1$, and the $P_{ij}$ are one-dimensional projectors.
We can then construct a new POVM by putting together all the
\mbox{$b_{ij}P_{ij}$} as elements.
It~is clear that these new elements are positive matrices and that
we still have $\sum_{ij} b_{ij}P_{ij} = \sum_i M_i = \I$.
To~obtain precisely the effect of the original POVM with the new one,
we must interpret the new POVM outcomes as follows:
If~the outcome $ij$ is obtained when the new POVM is applied,
we pretend that the outcome was simply~$i$.
Note that it could happen that $P_{ij}=P_{i^\prime j^\prime}$
for some $i^\prime \neq i$, but this does not cause an ambiguity
in the reinterpretation because the POVM outcome is actually $ij$,
not~$P_{ij}$.
\end{proof}

There are two natural and equivalent ways to represent projectors that act on single qubits.
As~a ket-bra, it is given by a matrix \mbox{$P=\ket{\Psi}\!\bra{\Psi}$}, for the
arbitrary one-qubit pure state $\ket{\Psi}$ on which projection is to be carried~out.
In~this case, it can always be rewritten in the form
\begin{equation}\label{projmat}
P=\left(
\begin{array}{cc}
\cos^2 \theta &  e^{-i\phi} \sin \theta \cos \theta  \\
e^{i\phi} \sin \theta \cos \theta   & \sin^2 \theta
\end{array}
\right)
\end{equation}
for appropriate angles
$0\leq \theta \leq \pi/2$ and  $0 \leq \phi \leq 2\pi$.
(Somewhat unconventionally, we~allow $\phi=2\pi$ for reasons that will soon be apparent.)

The same projector can also be represented as a three-dimensional unit vector
\begin{equation}\label{xyz}
\vec{v} = (x,y,z) = \bigl( \sin(2\theta) \cos(\phi), \sin(2\theta)  \sin(\phi), \cos(2\theta) \bigr)
\end{equation}
that can be seen as a point on the surface of a unit sphere (the Bloch Sphere) as follows.
Start with the vertical unit vector, which points to the north pole $(0,0,1)$, and make it tilt towards the front of the sphere\,%
\footnote{\,The tilt starts from the north pole towards the front, which is $(1,0,0)$, but if the angle of rotation
exceeds $\pi/2$, then the rotation continues downwards towards the south pole $(0,0,-1)$.}
by an angle of~$2\theta$.
Then make a sinistrorsal rotation about the vertical axis by an angle of~$\phi$.
We~say that the projector is in the \textit{east hemisphere} if $0 \le \phi < \pi$
and in the \textit{west hemisphere} if $\pi < \phi \le 2\pi$.
Note that a rotation of $\phi=2\pi$ has the same effect as no rotation at all ($\phi=0$), which means
that points with coordinates $(x,0,z)$ for positive $x$ belong to both hemispheres; we shall
make use of this apparent ambiguity later.
Note also that we have excluded $\phi=\pi$ from either hemisphere, which corresponds to points $(x,0,z)$ with negative~$x$.
The poles are singularities that deserve special treatment because when $\theta=0$ or $\theta=\pi/2$ the vector is
vertical after the tilt (remember that we tilt by angle $2\theta$), and therefore the rotation has no effect,
regardless of the value of~$\phi$.%
\footnote{\,Seen from the perspective of Equation~(\ref{projmat}) $\phi$ is irrelevant because
$\sin \theta \cos \theta = 0$ when $\theta=0$ or $\theta=\pi/2$.}
In~order to have a well-defined procedure in what follows, we stipulate that the north pole ($\theta=0$)
belongs to both hemispheres, whereas the south pole ($\theta=\pi/2$) belongs to neither.
To~enforce the latter condition, we declare that $\phi=\pi$ whenever $\theta=\pi/2$.
We~extend the notion of hemispheres to POVM elements proportional to projectors by saying
that $\gamma P$ belongs to the same hemisphere as~$P$, for any \mbox{$0 < \gamma \le 1$}.

The next lemma, lifted from~\cite{cgm99}, provides an alternative characterization of when
a collection of elements proportional to projectors forms a~POVM for the measurement of a single qubit.
It~is followed by our main technical lemma.

\begin{lemma}\label{cgm}
Consider a collection of projectors~$P_i$ and positive real numbers~$\gamma_i$.
For each~$i$, let $\vec{v}_i$ be the point on the Bloch Sphere that corresponds to~$P_i$,
according to Equation~(\ref{xyz}).
The~POVM condition $\sum_i \gamma_i P_i = \I$ is equivalent to saying that
\mbox{$\sum_i \gamma_i \vec{v}_i = 0$} and \mbox{$\sum_i \gamma_i = 2$}.
\end{lemma}

\begin{lemma}\label{hemisphere}
Any POVM whose elements are proportional to projectors contains at least one element
in the east hemisphere and at least one (possibly the same) in the west hemisphere.
\end{lemma}

\begin{proof}
Consider a POVM $\{ \gamma_i P_i \}$, where each $P_i$ is a projector and \mbox{$0 < \gamma_i \le 1$}.
For each $i$, let $\theta_i$, $\phi_i$ and \mbox{$\vec{v}_i=(x_i,y_i,z_i)$} correspond to~$P_i$
according to Equations~(\ref{projmat}) and~(\ref{xyz}).
If~at least one of the $P_i$ corresponds to the north pole (\mbox{$\theta_i=0$}),
or if some $\phi_i=0$ (equivalently $\phi_i=2\pi$),
this $\gamma_i P_i$ is a POVM element that belongs to both hemispheres.%
\footnote{\,This condition is equivalent to saying that $x_i \ge 0$, $y_i=0$ and $z_i \neq -1$.
For example, in the
case of a von Neumann measurement in the computational basis
\mbox{$\{\ket{0}\!\bra{0},\,\ket{1}\!\bra{1}\}$},
the $(0,0,1)$ vector,
which corresponds to seeing a 0 in the measurement apparatus,
belongs to both hemispheres, whereas the other vector, $(0,0,-1)$,
belongs to neither.}
Otherwise, the condition \mbox{$\sum_i \gamma_i \vec{v}_i = 0$} implies that there
must exist some $i$ such that \mbox{$y_i \neq 0$}.
If~\mbox{$y_i > 0$} (resp.~\mbox{$y_i < 0$}), then $\gamma_i P_i$ belongs to the east (resp.~west) hemisphere.
In~either case, we use condition \mbox{$\sum_i \gamma_i \vec{v}_i = 0$} again to conclude that there must be
some other projector $P_j$
such that the sign of $y_j$ is opposite to that of $y_i$, and therefore $\gamma_j P_j$ belongs to the other
hemisphere.
\end{proof}

Before we can proceed with the formal statement and proof of our main result,
we need a few additional technical lemmas.

\begin{lemma}\label{real}
Consider any two-player game that has a quantum winning strategy provided the
players share some state $\ket{\Phi}$ of dimension \mbox{$2 \times 2$}.
The same game also has a winning strategy if the players are
restricted to sharing a state of the form
\mbox{$\ket{\Psi}=\alpha\ket{00}+\beta\ket{11}$}, where
$\alpha$ and $\beta$ are well-chosen positive real numbers.
\end{lemma}
\begin{proof}
We~know from the Schmidt decomposition theorem that there exist
orthogonal bases \mbox{$\{\ket{A_0},\ket{A_1}\}$} for Alice and
\mbox{$\{\ket{B_0},\ket{B_1}\}$} for Bob such that $\ket{\Phi}$ can be rewritten as
\[ \ket{\Phi} = \alpha \ket{A_0}\ket{B_0} + \beta \ket{A_1}\ket{B_1} \]
for appropriate positive real numbers $\alpha$ and~$\beta$.
If~Alice and Bob share entangled state $\ket{\Psi}=\alpha\ket{00}+\beta\ket{11}$
instead of~$\ket{\Phi}$, Alice applies
unitary transformation \mbox{$\ket{A_0}\!\bra{0}+\ket{A_1}\!\bra{1}$} to her qubit
and Bob does the same with \mbox{$\ket{B_0}\!\bra{0}+\ket{B_1}\!\bra{1}$}.
The effect of those local quantum operations is to transform $\ket{\Psi}$ into~$\ket{\Phi}$.
From~there, Alice and Bob can apply the quantum strategy whose existence we assumed.
\end{proof}

\begin{lemma}\label{square}
Any two-party pseudo-telepathy game of dimension \mbox{$d_A \times d_B$} is also
a game of dimension \mbox{$d \times d$}, where \mbox{$d = \min(d_A,d_B)$}.
\end{lemma}
\begin{proof}
There is nothing to prove if \mbox{$d_A=d_B$}.
Assume without loss of generality that \mbox{$d = d_A < d_B$}.
We know from the Schmidt decomposition theorem that any
quantum \mbox{system} of dimension \mbox{$d_A \times d_B$} can be rewritten as
$\sum_i^d \alpha_i \ket{A_i}\ket{B_i}$
in appropriate orthogonal bases \smash{$\{\ket{A_i}\}_{i=1}^{d_A}$} for Alice
and \smash{$\{\ket{B_i}\}_{i=1}^{d_B}$} for Bob, where Bob's sub-basis $\{\ket{B_i}\}_{i=1}^d$
spans a \mbox{$d_A$-dimensional} subspace of his original $d_B$-dimensional
Hilbert space. From here, the proof follows
along the same lines as that of Lemma~\ref{real}.
\end{proof}

\begin{lemma}\label{moyennes} 
Consider any two positive numbers $a$ and $b$.
It is always the case that \mbox{$a^2+b^2 \ge 2ab$}, with equality holding
if and only if \mbox{$a=b$}.
\end{lemma}
\begin{proof}
The geometric average of $a^2$ and $b^2$ is \mbox{$\sqrt{a^2 b^2} = ab$} and their
arithmetic average is \mbox{$(a^2+b^2)/2$}.
The lemma follows from the well-known fact that the geometric average
of positive numbers is always a lower bound on their arithmetic average,
equality holding if and only if the numbers are equal.
\end{proof}

\section{Main result}\label{main}

We are now ready to state and prove our main result.

\begin{theo}\label{no2qubitspt}
There is no two-player pseudo-telepathy game of dimension \mbox{$2 \times 2$}.
\end{theo}
\begin{proof}

Consider any two-player game $(X,Y,A,B,P)$ for which Alice and Bob have a quantum winning strategy
in which they share a prior entangled state of dimension \mbox{$2 \times 2$}.
Our~goal is to exhibit a purely classical strategy that also wins the game.
By~definition, this means that the game in question is \textit{not} a pseudo-telepathy game,
whence the theorem is proven.

It~is important to understand that we
are \textit{not} trying to simulate the output \textit{probabilities} of the quantum strategy:
this would be impossible in general without communication due to Bell's theorem.
We~are not even trying to find a classical strategy that can produce with nonzero
probability exactly the set of outputs that the quantum strategy can produce with
nonzero probability: this would be equally impossible because of Hardy's state, as
explained at the end of the Introduction.
All~we are asking of our classical strategy is that it should \textit{never} produce
an illegal output even though some legal outputs may never occur.
This condition will be automatically fulfilled if we design our classical strategy
in a way that it will never produce an output that would have had zero probability
of being produced by the quantum strategy, since we are assuming that the quantum
strategy wins the game.

According to quantum mechanics, the most general strategy that Alice and Bob can deploy
consists in each of them independently choosing a POVM depending on their inputs,
applying that POVM on their share of the entanglement, and interpreting the outcome of
their measurements in terms of elements of their output sets.
More formally, let $\mathcal{P}$ denote the set of all POVMs acting on a single qubit.
For each $M \in \mathcal{P}$, let $\{M_i\}$ denote the corresponding set of positive matrices,
with \mbox{$\sum_i M_i = \I$} of course, and let $O_M$ denote the set of possible outcomes
for that POVM, i.e.~the index $i$ in $\{M_i\}$ ranges over all the values in~$O_M$.
Let~$O$ denote the union of all~$O_M$ for $M \in \mathcal{P}$.%
\footnote{\,Of course, $O$ is simply the set of all natural numbers, but it is better for
the intuition to think of it as the set of all possible measurement outcomes.}

Any quantum strategy can be defined in terms of the shared quantum state $\ket{\Psi}$
and the following mappings.
\[
\begin{array}{ll}
\mathcal{X} ~:~ X \rightarrow \mathcal{P} & ;~~~\mathcal{Y} ~:~ Y \rightarrow \mathcal{P} \\
\mathcal{A} ~:~ X \times O \rightarrow A &  ;~~~\mathcal{B} ~:~ Y \times O \rightarrow B
\end{array}
\]
Upon receiving her input $x \in X$, Alice determines her measurement $M^x = \mathcal{X}(x)$ and
applies it to her share of~$\ket{\Psi}$.  She obtains some outcome \mbox{$i \in O_{M^x}$}.
From this, she outputs $\mathcal{A}(x,i)$.
Upon receiving his input \mbox{$y \in Y$}\!, Bob does the same, \textit{mutatis mutandis}.
This is truly the most general form of quantum strategy, since all one can do with a
quantum system is add an ancillary system, do a unitary
transformation and perform a von Neumann measurement, all of which is
covered by the POVM formalism, thanks to Naimark's theorem.

Without loss of generality, according to Lemma~\ref{real}, we may assume that
Alice and Bob's winning strategy uses an entangled state
of the form $\ket{\Psi}=\alpha\ket{00}+\beta\ket{11}$, where
$\alpha$ and $\beta$ are positive real numbers.
We~may as well assume that $\alpha$ and $\beta$ are in fact \textit{strictly} positive
because otherwise $\ket{\Psi}$ is a product state and local measurements on pure states would be
easy to simulate classically by Alice and Bob.
Also without loss of generality, according to Lemma~\ref{sd}, we may assume that the
POVM elements that appear in the image of $\mathcal{X}$ and~$\mathcal{Y}$ are proportional
to projectors.

Consider an instance of the game in which Alice and Bob receive inputs \mbox{$x \in X$} and
\mbox{$y \in Y$}\!, respectively.
According to the quantum strategy,
let \mbox{$M^x=\mathcal{X}(x)=\{\gamma_i^x P_i^x\}$}
and \mbox{$N^y=\mathcal{Y}(y)=\{\gamma_j^y Q_j^y\}$}
be the POVMs applied by Alice and Bob%
\footnote{\,Formally speaking, we should write $\mbox{}^{A} \gamma_i^x$ and $\mbox{}^{B} \gamma_j^y$
to distinguish the $\gamma$s of Alice from those of Bob, or else use different Greek letters,
because it could happen that Alice and Bob are given the same input (\mbox{$x=y$}), yet the
$\gamma^x$ and $\gamma^y$ are different altogether.
To avoid cluttering the notation, however,
it~will be implicitly understood that the superscripts $x$ and $y$ serve also as labels that identify
ownership by Alice or Bob, in addition to their specific values as inputs from $X$ and~$Y$.
The same remark applies to other Greek letters such as angles $\theta$ and~$\phi$.},
respectively, on their share of the entanglement.
Let $\theta_i^x$ and $\phi_i^x$ be the angles that characterize projector $P_i^x$
according to Equation~(\ref{projmat}), and similarly for $\theta_j^y$ and $\phi_j^y$.
For any given $i$ and $j$, what is the joint probability $\Pr[i,j]$ that the outcome of Alice's
measurement be $i$ and of Bob's measurement be~$j$, simultaneously?
A~straightforward but tedious calculation yields the following:
\begin{equation}
\label{probs}
\begin{split}
\Pr[i,j] = & \, \bra{\Psi} (\gamma_i^x P_i^x) \otimes (\gamma_j^y Q_j^y) \ket{\Psi}\\[1ex]
= & \, \gamma_i^x \gamma_j^y \bigl[ \alpha^2 \cos(\theta_i^x)^2 \cos(\theta_j^y)^2 +
\beta^2 \sin(\theta_i^x)^2 \sin(\theta_j^y)^2 \\
& \phantom{\, \gamma_i^x \gamma_j^y \bigl[} +2 \alpha \beta \cos(\theta_i^x) \cos(\theta_j^y) \cos(\phi_i^x +
\phi_j^y) \sin(\theta_i^x) \sin(\theta_j^y) \bigr].
\end{split}
\end{equation}
Let \mbox{$a= \alpha \cos(\theta_i^x) \cos(\theta_j^y)$},
\mbox{$b=\beta \sin(\theta_i^x) \sin(\theta_j^y)$} and \mbox{$c=\cos(\phi_i^x+\phi_j^y)$}.
Note that $a \ge 0$ and $b \ge 0$ because $\alpha>0$, $\beta>0$,
\mbox{$0 \le \theta_i^x \le \pi/2$} and \mbox{$0 \le \theta_j^y \le \pi/2$},
and of course \mbox{$-1 \le c \le 1$}.
Therefore, because $\gamma_i^x$ and $\gamma_j^y$ are nonzero, it follows from Lemma~\ref{moyennes} that
the joint probability \mbox{$\Pr[i,j] = \gamma_i^x \gamma_j^y (a^2 + b^2 + 2 a b c)$}
can only vanish if \mbox{$a=b=0$} or if \mbox{$a=b$} and \mbox{$c=-1$}.
The first case requires that $\theta_i^x=0$ and $\theta_j^y=\pi/2$ or \textit{vice versa},
which means that one of $P_i^x$ or $Q_j^y$ belongs to neither hemisphere (being proportional to the south pole).
The~condition $c=-1$ in the second case implies that \mbox{$\phi_i^x+\phi_j^y=\pi$}
or \mbox{$\phi_i^x+\phi_j^y=3\pi$} because \mbox{$0 \le \phi_i^x+\phi_j^y \le 4\pi$}.

Recall that our purpose is to determine a classical strategy between Alice and Bob
that will never produce a joint output whose probability would have vanished according
to the quantum winning strategy.  To~achieve this goal, it suffices for Alice
to select an $i$ such that $\gamma_i^x P_i^x$ belongs to the east hemisphere
and for Bob to select a $j$ such that $\gamma_j^y Q_j^y$ belongs to the west hemisphere
(without actually measuring anything).
This is always possible according to Lemma~\ref{hemisphere}.
In~this way, neither Alice nor Bob will choose a POVM element proportional to the south pole
(thus avoiding \mbox{$a=b=0$}),
and \mbox{$\pi < \phi_i^x+\phi_j^y < 3\pi$} since
\mbox{$0 \le \phi_i^x < \pi$} and \mbox{$\pi < \phi_j^y \le 2\pi$}
(thus avoiding \mbox{$c=-1$}).
It~follows that the choices made independently by Alice and Bob with this
classical strategy correspond to choices that Alice and Bob could have made with
nonzero probability had they followed the quantum strategy.
By~assumption that the quantum strategy would have produced a valid response,
so is the case if the classical strategy is to output $\mathcal{A}(x,i)$ for Alice
and $\mathcal{B}(y,j)$ for~Bob.
\end{proof}

\begin{corol}\label{cor}
There is no two-player pseudo-telepathy game of dimension $2 \times n$,
no matter the value of integer~$n$.
\end{corol}
\begin{proof}
According to Lemma~\ref{square}, any pseudo-telepathy game of dimension \mbox{$2 \times n$}
would also be a pseudo-telepathy game of dimension \mbox{$2 \times 2$}.
But according to Theorem~\ref{no2qubitspt}, no such game can exist.
\end{proof}

\begin{corol}\label{cortwo}
The optimal two-player pseudo-telepathy game requires a dimension \mbox{$3 \times 3$}.
\end{corol}
\begin{proof}
This is immediate from Corollary~\ref{cor} and the fact that a
pseudo-telepathy game of dimension \mbox{$3 \times 3$}
is known to exist~\cite{chtw04}.
\end{proof}

\begin{corol}
The optimal pseudo-telepathy game, in terms of the total dimension of the required entangled state,
is a \textit{three}-player game of dimension~\mbox{$2 \times 2 \times 2$}.
\end{corol}
\begin{proof}
A~meaningful pseudo-telepathy game requires each player to have at least one qubit of the shared entanglement.
It~follows that an $n$-player pseudo-telepathy game must be of total dimension at least~$2^n$ if all players are
to participate quantum mechanically in the game.
According to Corollary~\ref{cortwo}, the best two-player pseudo-telepathy game is of dimension \mbox{$3 \times 3 = 9$}.
According to the discussion above, the three-player GHZ/Mermin pseudo-telepathy game~\cite{ghz89,mermin90a},
which is of dimension \mbox{$2 \times 2 \times 2 = 8$},
is optimal among three-player pseudo-telepathy games.
Adding more players would only increase the dimension to at least \mbox{$2^n \ge 16$} for \mbox{$n \ge 4$} players.
The corollary follows from the fact that \mbox{$8 < 9 < 16$}.
\end{proof}

\section{Conclusion}\label{concl}

In conclusion, we have proven that the pseudo-telepathy game of CHTW~\cite{chtw04},
which uses two entangled qutrits and outputs a bit and a trit, is the minimal
possible two-player pseudo-telepathy game.
Nevertheless, in
terms of the total dimension of the composite quantum system,
this two-player game is beaten by Mermin's three-player
pseudo-telepathy game~\cite{mermin90a}.

\newpage

The technique used by Aravind~\cite{aravind99} and CHTW~\cite{chtw04} to
build a pseudo-telepathy game of dimension~\mbox{$d \times d$} from any
Kochen-Specker construction of dimension~$d$ does not require Alice and Bob to
perform generalized measurements on their quantum systems.
It~is tempting to think that this comes from the fact that the standard Kochen-Specker theorem
is stated in terms of projective measurements~\cite{KS67}.
However, it has been suggested that the Kochen-Specker theorem could be
extended by using POVMs~\cite{cabello02,toner}.
In~particular, this makes it possible to consider a Kochen-Specker theorem for a single qubit,
which would be obviously impossible with the standard approach.
Could the technique of Aravind and CHTW extend
to those POVM-based Kochen-Specker theorems and yield pseudo-telepathy games of the same dimension,
except that Alice and Bob would have to perform POVMs on their share of the entanglement?
Unfortunately, our result implies that such hopes are doomed because
there \textit{is} a POVM-based Kochen-Specker theorem of dimension~2, but there
can\textit{not} be a pseudo-telepathy game of dimension \mbox{$2 \times 2$}, even if POVMs
are used.

In this paper, we have established that POVMs confer no advantage to pseudo-telepathy
strategies, compared to simpler projective von Neumann measurements,
when the quantum system shared between Alice and Bob is restricted to being
of dimension~\mbox{$2 \times 2$}.
What is the situation in higher dimensions or with more players?
Can~\textit{any} pseudo-telepathy game of dimension~\mbox{$d \times d$} be won
with a strategy of the same dimension in which the players perform only
projective measurements?
A~figure of merit for any given pseudo-telepathy game is the best success probability
possible by any purely classical strategy~\cite{BBT04a}.
The smaller is this probability,
the more difficult is the game classically, and therefore the more surprised
a classical physicist would become at the systematic success of our quantum players.
This probability must be strictly smaller than~1 by definition of pseudo-telepathy, but some games
are known for which it is almost ridiculously close to~1~\cite{gwt02}.
For~any positive integer $d$, one can consider the success probability $p_d$
of the best classical algorithm for the classically most challenging pseudo-telepathy game
of dimension~\mbox{$d \times d$}.
Can~$p_d$ be smaller (i.e.~better) if we allow quantum strategies that
use POVMs, rather than restricting Alice and Bob to performing only projective measurements?
Is~the situation different for multi-player games? 

\section*{Acknowledgements}

The authors are grateful to Anne Broadbent and Serge Massar for helpful comments.

\end{document}